\documentclass{SCIS2015author}
\usepackage{amsmath}
\begin{document}

%%%%%%%%%%%%%%%%%%%%%%%%%%%%%%%%%%%%%%%%%%%%%%%%%%%%%%%
%%% Authors do not modify the information below
%%% 作者不需要修改此处信息
%%% 有专题名称时, 将第一行的{}注释掉, 使用第二行
\ArticleType{RESEARCH PAPER}{}
%{\put(0,-26){专题名称}}
\Year{2015}
\Month{June}
\Vol{xx}
\No{xx}
\DOI{xxxxxxxxxxxxxx}
\ArtNo{xx}
\ReceiveDate{xxxx, 2015}
\AcceptDate{xxxx, 2015}
\OnlineDate{xxxx, 2015}
%%%%%%%%%%%%%%%%%%%%%%%%%%%%%%%%%%%%%%%%%%%%%%%%%%%%%%%

%%% title: 标题
%%%   \title{full title}{title for citation}
\title{Transceiver designs with matrix-version water-filling architecture under mixed power constraints}{Transceiver design with matrix-version water-filling architecture with mixed power constraints}

%%% Corresponding author: 通信作者
   %%%\author[1]{Xing C W}{{xingchengwen@gmail.com}}
%%% General author: 一般作者
%%%   \author[number]{Full name}{}
\author[1]{XING ChengWen}{{xingchengwen@ieee.org}}
\author[1]{FEI ZeSong}{}
\author[2]{ZHOU YiQing}{}
\author[3]{\\PAN ZhenGang}{}
\author[3]{WANG HuaLei}{}

%%% Author information for page head. 页眉中的作者信息
\AuthorMark{Xing C W}

%%% Authors for citation. 首页引用中的作者信息
\AuthorCitation{Xing C W, Fei Z S, Zhou Y Q, et al}

%%% Address: 地址
%%%   \address[number]{Address, City {\rm Postcode}, Country}
\address[1]{School of Information and Electronics, Beijing Institute of Technology,\\ Beijing. {\rm 100081}, China}
\address[2]{Wireless Communication Research Center,
Institute of Computing Technology, \\Chinese Academy of Sciences, Beijing, {\rm 100190}, China}
\address[3]{China Mobile Research Institute,\\ Beijing {\rm 100053}, China}
\maketitle

%%% Abstract: 摘要
%%% Not necessery for LETTER. LETTER不需要摘要
\abstract{In this paper, we investigate the multiple-input multiple-output (MIMO) transceiver design under an interesting power model named mixed power constraints. In the considered power model, several antenna subsets are constrained by sum power constraints while the other antennas are subject to per-antenna power constraints. This kind of transceiver designs includes both the transceiver designs under sum power constraint and per-antenna power constraint as its special cases. This kind of designs is of critical importance for distributed antenna systems (DASs) with heterogeneous remote radio heads (RRHs) such as cloud radio access networks (C-RANs). In our work, we try to solve the optimization problem in an analytical way instead of using some famous software packages e.g., CVX or SeDuMi. In our work, to strike tradeoffs between performance and complexity, both iterative and non-iterative solutions are proposed. Interestingly the non-iterative solution can be interpreted as a matrix version water-filling solution extended from the well-known and extensively studied vector version. Finally,  simulation results demonstrate the accuracy of our theoretical results.}

%%% Keywords: 关键词
%%% Not necessery for LETTER. LETTER不需要关键词
\keywords{Convex optimization, MIMO, matrix-version water-filling, transceiver designs, mix power constraints.}

\citationline

%%%%%%%%%%%%%%%%%%%%%%%%%%%%%%%%%%%%%%%%%%%%%%%%%%%%%%%
%%% The main text. 正文部分
%%% No section name for LETTER. LETTER不分章节
%%%%%%%%%%%%%%%%%%%%%%%%%%%%%%%%%%%%%%%%%%%%%%%%%%%%%%%
\section{Introduction}

With channel state information (CSI), multiple-input multiple-output (MIMO) transceiver designs can significantly improve the performance of the considered communication systems \cite{Telatar1999,Palomar03,Scaglione2002,JYangh1994,Sampth01}. For transceiver designs, power constraints are the most natural and fundamental constraints that should be carefully addressed. With an antenna array deployed at transmitter, the most widely used power constraint is sum power constraint i.e., the sum of transmit powers at different antennas is smaller than a threshold \cite{Palomar03,Telatar1999,Sampth01}. Later individual power constraint or per-antenna power constraint has been recognized as a more practical one than sum power constraint. The reason is that in practice each antenna has its own amplifier and is limited individually by its own amplifier's maximum power threshold \cite{YUWEI2007,Tolli2008,huilin_zhu_2}.

An interesting question is whether the sum power constraint is impractical and meaningless. The answer is definitely ``not''. Although different antennas have their own individual amplifiers, it has been shown in \cite{Mai2011} that when each amplifier has the same power constraint, the gap between the designs with individual power constraint and sum power constraint is very small and almost negligible. Importantly, the design under sum power constraint is much easier than its counterpart under individual power constraints \cite{Tolli2008,Huiling_Zhu_Resource,XinChen2012}. It can be concluded that if the wireless terminal is equipped with the same style antennas sum power constraint is a reasonable assumption in the sense of engineering designs and this may be the reason why sum power constraints are always chosen for transceiver designs \cite{Palomar03,Telatar1999,Sampth01}.

On the other hand, can we say individual power constraints are less important. The answer is also ``not''. For some special network architectures, e.g., distributed antenna systems (DASs) \cite{Huiling_Zhu_Resource}, the largely separated distributed antennas can have different sizes and then to avoid a drastic performance degradation, individual power constraints cannot be replaced by a simple sum power constraint \cite{huilin_zhu_2,Chen2010,Gao1}. Regarding this fact, individual power constraints (per-antenna power constraints) have attracted a lot of attention from wireless researchers in various areas \cite{FeifeiGao,NaLiCL,Wang2012,Chen2010}. However the transceiver designs under individual power constraints are more challenging than their counterparts under sum power constraint, in most cases the design problems can still be cast into standard convex optimization problems such Semi-definite programming (SDP), Second order Cone Programming (SOCP), etc. Then they can be efficiently solved by using some well-known software toolboxes  \cite{Chen2010}. In some special cases, relying on Lagrange dual functions, the considered optimization problems can be efficiently solved by sub-gradient methods \cite{NaLiCL}.

Both sum power constraint and individual power constraint are important for wireless designs. The design with sum power constraint is much simpler as it has less constraints and then less Lagrange multipliers to compute.  On the other hand, the design with individual power constraint is more realistic and usually has better performance for DASs especially when antennas are powered by amplifiers with different specifications.  It should  be highlighted that sum power constraint cannot be considered as a special case of individual power constraint although individual power constraint is much stricter.
In this paper, we take a further step to investigate a more general case named mixed power constraint. Under mixed power constraint, for a given set of antennas, several subsets of the antennas are subject to sum power constraints and the other antennas have their own individual power constraints. This design can realize a tradeoff between sum power constraint and individual power constraint/per-antenna power constraint. In our design,  the antennas of the same size at a certain wireless terminal can be taken as a cluster under a sum power constraint at the expense of a slight performance loss. Then the corresponding computation procedure is significantly simplified.

Actually, this kind of power constraint can be applied to several practical scenarios. For example, in a cloud radio access network (C-RAN), radio units are separated from baseband units (BBUs) and installed distributively. To realize a dense and seamless coverage, the remote radio heads (RRHs) may have different settings, e.g., different sizes, different antenna elements and so on \cite{CRAN}. For each RRH, more than one antenna can be installed and these antennas have the same specifications. Therefore at each RRH the power constraint can be modeled as sum power constraint with slight or even negligible performance loss, while among different RRHs it is natural to choose individual power constraints to avoid significant performance losses.

A natural question is how to design the transceivers under mixed power constraints, numerically or analytically?
Recently, an interesting and excellent work shows that even under individual power constraints, an analytical iterative method can be used to give satisfied solutions \cite{Mai2011}.  This work motivates the authors to think whether this logic can be extended to the transceiver design with mixed power constraints and the final solutions can be derived in much simple closed forms without too many mathematical symbols. Complicated mathematics can reveal some important performance bounds, however too many mathematical symbols and substitutions in the formulation will bury its physical meanings and prohibit the practical implementations. For engineers, the solutions should be as simple as possible. Motivated by these facts, we try to solve the optimization problem in a much simpler manner. The main contributions of this work are listed as follows.

\noindent $\bullet$ Firstly, in our work we investigate the transceiver design under mixed power constraint which can realize any tradeoff between sum power constraint and per-antenna power constraint. From the theoretical viewpoint mixed power constraint includes both sum power constraint and per-antenna power constraint as its special cases. To the best of our knowledge, it is the first time to investigate the transceiver designs for MIMO systems under mixed power constraints.

\noindent $\bullet$ Secondly, the formula of the optimal solution of the transceiver design under mixed power constraint has been derived. Here, we would like to highlight that the derivation logic of our work is significantly different from that in \cite{Sampth01,Mai2011}. In the special case with only per-antenna power constraints, compared to \cite{Mai2011} our derivation procedure does not need case by case discussions i.e., classifying channel matrices into tall matrices or fat matrices.

\noindent $\bullet$ Thirdly, in our work except an iterative solution, a non-iterative solution is proposed as well, which does not need iterations and can be interpreted as a matrix version water-filling solution. It is shown by simulation results that the non-iterative solution has a satisfied performance.

\noindent $\bullet$ Finally, for our work there is no constraint on the rank of channel matrix, however in the existing work \cite{Mai2011} the channel matrix should be column full rank or row full rank.

\noindent \textbf{Notation:} Throughout this paper, the following notations are used. First, boldface
lowercase letters denote vectors and boldface uppercase letters
denote matrices, respectively.
Transpose and Hermitian transpose of a matrix are denoted by
$(\cdot)^{\rm{T}}$ and $(\cdot)^{\rm{H}}$. $\rm{Tr}\{\cdot\}$ and
$\rm{rank}(\cdot)$ are used to represent the trace and the rank of a
matrix, respectively. The symbol $\rm{\mathbb{E}}\{\cdot\}$
represents the expectation operation. ${\rm{I}}_M$ denotes the
$M\times M$ identity matrix.
In addition, ${\rm{Tr}}({\bf{Z}})$ is the
trace of the matrix ${\bf{Z}}$. The notation ${\bf{Z}}^{1/2}$ is the
Hermitian square root of the positive semidefinite matrix
${\bf{Z}}$. The symbol $[{\bf{Z}}]_{i,j}$ represents the $\{i,j\}^{\rm{th}}$ element of the matrix ${\bf{Z}}$. For two Hermitian matrices, ${\bf{C}} \succeq
{\bf{D}}$ means that ${\bf{C}}-{\bf{D}}$ is a positive semi-definite
matrix.

\section{System Model and Problem Formulation}

In this work, we consider a simple point-to-point MIMO system. The received signal equals ${\bf{y}}={\bf{H}}{\bf{x}}+{\bf{n}}$ where ${\bf{x}}$ is the transmitted signal from the source with covariance matrix ${\bf{Q}}$ and ${\bf{H}}$ is the MIMO channel matrix. In addition, ${\bf{n}}$ denotes additive Gaussian noise and the noise covariance matrix is ${\bf{R}}_n$. For MIMO systems, transmitter at source node can be understood as a spatial filter and optimizing transmitter is equivalent to optimizing ${\bf{Q}}$. For the system model ${\bf{y}}={\bf{H}}{\bf{x}}+{\bf{n}}$ with Gaussian distributed noises, the optimal receiver is ${\bf{G}}={\bf{Q}}{\bf{H}}^{\rm{H}}({\bf{H}}{\bf{Q}}{\bf{H}}^{\rm{H}}+
{\bf{R}}_{\bf{n}})^{-1}$ \cite{Palomar03}, which is also a function of ${\bf{Q}}$. Therefore, the considered transceiver optimization problem becomes how to optimize ${\bf{Q}}$.

Based on the discussions above, the capacity maximization problem for a point-to-point MIMO system under mixed power constraint is formulated as
\begin{align}
\label{Matrix_Opt}
& \max \ \ {\rm{log}}|{\bf{I}}
+{\bf{H}}{\bf{Q}}{\bf{H}}^{\rm{H}}{\bf{R}}_{{\bf{n}}}^{-1}|\nonumber \\
& \ {\rm{s.t.}} \ \ \ {\sum}_{j\in \psi_k}\{{\bf{e}}_j^{\rm{H}}{\bf{Q}}{\bf{e}}_j\}
\le p_k, \nonumber \\
&\ \ \ \ \ \ \ \ {\bf{Q}} \succeq {\bf{0}}.
\end{align}where ${\bf{Q}}$ denotes the covariance matrix of the transmitted signal  and ${\bf{H}}$ is the MIMO channel matrix. In addition, the noise covariance matrix is ${\bf{R}}_n$. The symbol ${\bf{e}}_j$ denotes the vector with the $j^{\rm{th}}$ element being 1 and the other elements being zeros.
Furthermore, $\psi_k$ is a subset of the antenna index set $\{1,2,\cdots, N\}$, and notice that for $k_1 \not= k_2$, $\psi_{k_1} \cap \psi_{k_2}=\emptyset$ and ${\cup}\psi_{k}=\{1,2,\cdots, N\}$. The antennas in the same $\psi_k$ are subject to a certain sum power constraint.

Due to various antenna settings, i.e., the number of transmit antennas is larger or smaller than that of receiver antennas, the channel matrix ${\bf{H}}$ may be a tall or fat matrix. To avoid case-by-case discussions, exploiting the fact that the considered objective function is continuous an auxiliary variable $\alpha$ is introduced first and then the following optimization problem will have the same optimal solution as that of the original optimization problem (\ref{Matrix_Opt})
\begin{align}
\label{matrix_optimization_0}
& \max_{{\bf{Q}}_k} \ \ \lim_{\alpha \rightarrow 0} {\rm{log}}|{\bf{I}}
+({\bf{H}}^{\rm{H}}{\bf{R}}_{{\bf{n}}}^{-1}
{\bf{H}}+\alpha{\bf{I}}){\bf{Q}}|\nonumber \\
& \ \ {\rm{s.t.}} \ \ {\sum}_{j\in \psi_k}\{{\bf{e}}_j^{\rm{H}}{\bf{Q}}{\bf{e}}_j\} \le p_k \nonumber  \\
& \ \ \ \ \ \ \ \ {\bf{Q}} \succeq {\bf{0}}.
\end{align} Inspired by this fact, in the next section we will concentrate our attention on the following optimization problem whose optimal solution is in nature a function of $\alpha$ and then we should take the limit $\alpha \rightarrow 0$ on the derived optimal solution in order to achieve the exact optimal solution
\begin{align}
\label{matrix_optimization}
& \max_{{\bf{Q}}} \ \ {\rm{log}}|{\bf{I}}
+({\bf{H}}^{\rm{H}}{\bf{R}}_{{\bf{n}}}^{-1}{\bf{H}}
+\alpha{\bf{I}}){\bf{Q}}| \nonumber \\
& \ \ {\rm{s.t.}} \ {\sum}_{j\in \psi_k}\{{\bf{e}}_j^{\rm{H}}{\bf{Q}}{\bf{e}}_j\} \le p_k \nonumber \\
& \ \ \ \ \ \ \   {\bf{Q}} \succeq {\bf{0}}.
\end{align}

\section{Matrix Version Water-Filling Solution}

Introducing the auxiliary variables $d_k$'s as the corresponding Lagrange multipliers to the $k$ power constraints, the Lagrange function of the optimization problem (\ref{matrix_optimization}) is given as
\begin{align}
\label{Lagrange_Dual}
{\mathcal{L}}({\bf{Q}},\{d_k\},{\boldsymbol \Psi})
=&{\rm{log}}|{\bf{I}}
+({\bf{H}}^{\rm{H}}{\bf{R}}_{{\bf{n}}}^{-1}{\bf{H}}+\alpha{\bf{I}}){\bf{Q}}|+\sum_k\{d_k[p_k-
\sum_{j\in \psi_k}\{{\bf{e}}_j^{\rm{H}}
{\bf{Q}}{\bf{e}}_j\}]\}+{\rm{Tr}}({\boldsymbol \Psi}{\bf{Q}})\nonumber \\
=&{\rm{log}}|{\bf{I}}
+({\bf{H}}^{\rm{H}}{\bf{R}}_{{\bf{n}}}^{-1}{\bf{H}}
+\alpha{\bf{I}}){\bf{Q}}| +\sum_k\{d_kp_k\}
-\sum_k\{{\rm{Tr}}({\bf{D}}{\bf{Q}})\}+{\rm{Tr}}({\boldsymbol \Psi}{\bf{Q}})
\end{align}where ${\bf{D}}$ is a diagonal matrix defined as
\begin{align}
\label{D}
[{\bf{D}}]_{j,j}=d_k \ \  \ j\in \psi_k.
\end{align}Notice that the diagonal elements of ${\bf{D}}$, ${d}_{k}$'s, must be positive. Additionally, the Hermitian matrix ${\boldsymbol \Psi}$ is the corresponding Lagrange multiplier of the constraint $ {\bf{Q}} \succeq {\bf{0}}$.    Based on convex optimization theory, if
$d_k=0$ it means that its corresponding power constraint is inactive and in other words this constraint can be simply removed from the considered optimization problem. For a practical wireless system, with proper interference mitigation, increasing transmit power seems always beneficial to the whole performance (at least not harmful). In this paper, we only consider active power constraints.

For the simplicity of exposition, defining a new matrix given as ${\boldsymbol{\mathcal{H}}} \triangleq  ({\bf{H}}^{\rm{H}}{\bf{R}}_{{\bf{n}}}^{-1}{\bf{H}}+\alpha{\bf{I}})^{1/2} $ and $ {\boldsymbol{\mathcal{H}}}={\boldsymbol{\mathcal{H}}}^{\rm{H}}
$ and substituting ${\boldsymbol{\mathcal{H}}}$ into (\ref{Lagrange_Dual}), the Karush-Kuhn-Tucker (KKT) conditions of (\ref{matrix_optimization}) can be derived to be
\begin{align}
\label{KKT_Matrix}
& {\boldsymbol{\mathcal{H}}}^{\rm{H}}({\bf{I}}
+{\boldsymbol{\mathcal{H}}}
{\bf{Q}}{\boldsymbol{\mathcal{H}}}^{\rm{H}})^{-1}
{\boldsymbol{\mathcal{H}}}
={\bf{D}}-{\boldsymbol \Psi} \nonumber \\
& d_k[p_k-{\rm{Tr}}(\sum_j\{{{\bf{e}}_j\bf{e}}_j^{\rm{H}}\} {\bf{Q}})]=0 \nonumber \\
& d_k \ge 0, \ \ {\rm{Tr}}({\bf{Q}}{\boldsymbol{\Psi}})=0, \ \ {\boldsymbol{\Psi}} \succeq {\bf{0}} \nonumber \\
& \sum_{j\in \psi_k}\{{\bf{e}}_j^{\rm{H}}{\bf{Q}}{\bf{e}}_j\} \le p_k, \ \  {\bf{Q}}\succeq {\bf{0}},
\end{align}where the first KKT condition comes from $\partial{\mathcal{L}}({\bf{Q}},\{d_k\},{\boldsymbol \Psi})/\partial {\bf{Q}}=0$. As the optimization problem (\ref{Matrix_Opt}) is convex, the KKT conditions are the necessary and sufficient conditions for optimal solutions \cite{Boyd04}. Based on the first KKT condition in (\ref{KKT_Matrix}), the covariance matrix ${\bf{Q}}$
 can be solved to be\begin{align}
\label{equ_Q}
{\bf{Q}}&=({\bf{D}}-{\boldsymbol \Psi})^{-1}-({\boldsymbol{\mathcal{H}}}^{\rm{H}}
{\boldsymbol{\mathcal{H}}})^{-1}\nonumber \\
&=[{\bf{D}}^{1/2}({\bf{I}}-{\bf{D}}^{-1/2}{\boldsymbol \Psi}{\bf{D}}^{-1/2}){\bf{D}}^{1/2}]^{-1}-({\boldsymbol{\mathcal{H}}}^{\rm{H}}
{\boldsymbol{\mathcal{H}}})^{-1}\nonumber \\
&={\bf{D}}^{-\frac{1}{2}}[({\bf{I}}-{\bf{D}}^{-\frac{1}{2}}
{\boldsymbol \Psi}{\bf{D}}^{-\frac{1}{2}})^{-1} ]{\bf{D}}^{-\frac{1}{2}}-({\boldsymbol{\mathcal{H}}}^{\rm{H}}
{\boldsymbol{\mathcal{H}}})^{-1}.
\end{align}Note that as the term $({\bf{I}}-{\bf{D}}^{-1/2}{\boldsymbol \Psi}{\bf{D}}^{-1/2})^{-1}$ has a complicated mathematical formula for further analysis and derivation, introducing an auxiliary matrix ${\boldsymbol \Pi}$ we will use a new formulation of $({\bf{I}}-{\bf{D}}^{-1/2}{\boldsymbol \Psi}{\bf{D}}^{-1/2})^{-1}$ for further analysis, i.e.,
\begin{align}
\label{simplify_equ}
({\bf{I}}-{\bf{D}}^{-1/2}{\boldsymbol \Psi}{\bf{D}}^{-1/2})^{-1}\triangleq {\bf{I}}+{\bf{D}}^{1/2}{\boldsymbol \Pi}{\bf{D}}^{1/2},
\end{align}where ${\boldsymbol \Pi}$ is a positive semi-definite matrix that replaces the role of  ${\boldsymbol \Psi}$.
Substituting (\ref{simplify_equ}) into (\ref{equ_Q}), we directly have
\begin{align}
\label{Q_3}
{\bf{Q}}&={\bf{D}}^{-1} -({\boldsymbol{\mathcal{H}}}^{\rm{H}}
{\boldsymbol{\mathcal{H}}})^{-1}
+{\boldsymbol \Pi},
\end{align}based on which it can be seen that the term ${\boldsymbol \Pi}$ guarantees that ${\bf{Q}}$ is positive semi-definite. Then the problem we are faced with is what characteristics ${\boldsymbol\Pi}$ owns can guarantee a suitable ${\boldsymbol\Psi}$ can be computed.

\noindent \textbf{Conclusion 1:} In order to find a suitable ${\boldsymbol\Psi}$,
${\boldsymbol\Pi}$ must satisfy
\begin{align}
{\rm{Tr}}[({\bf{I}} -{\bf{D}}^{1/2}
({\boldsymbol{\mathcal{H}}}^{\rm{H}}
{\boldsymbol{\mathcal{H}}})^{-1}
{\bf{D}}^{1/2}+{\bf{D}}^{1/2}{\boldsymbol \Pi}{\bf{D}}^{1/2}){\bf{D}}^{1/2}{\boldsymbol \Pi}{\bf{D}}^{1/2}]=0.
\end{align}
\begin{proof}
See the Appendix in~\ref{App_1}.
\end{proof}

Based on the following eigenvalue decomposition (EVD) with eigenvalues in decreasing order
\begin{align}
\label{equ_Q1}
{\bf{I}} -{\bf{D}}^{1/2}({\boldsymbol{\mathcal{H}}}^{\rm{H}}
{\boldsymbol{\mathcal{H}}})^{-1}{\bf{D}}^{1/2}=
{\bf{U}}_{{\bf{M}}}
{\boldsymbol \Lambda}_{{\bf{M}}}{\bf{U}}_{{\bf{M}}}^{\rm{H}},
\end{align} Conclusion 1 will be satisfied if the following equality holds
\begin{align}
\label{Psi_3}
{\bf{D}}^{1/2}{\boldsymbol \Pi}{\bf{D}}^{1/2}={\bf{U}}_{{\bf{M}}}
{\boldsymbol \Lambda}_{{\bf{M}}}^{-}
{\bf{U}}_{{\bf{M}}}^{\rm{H}},
\end{align}where the symbol $-$ is defined as for a diagonal matrix ${\bf{Z}}$ with ${\bf{X}}=[{\bf{Z}}]^{-}$ if $[{\bf{Z}}]_{i,i}\ge 0$, $[{\bf{X}}]_{i,i}=0$ and otherwise $[{\bf{X}}]_{i,i}=-[{\bf{Z}}]_{i,i}$.
It is obvious that the term ${\boldsymbol \Pi}$ in (\ref{Q_3}) guarantees ${\bf{D}}^{-1} -({\boldsymbol{\mathcal{H}}}^{\rm{H}}{\boldsymbol{\mathcal{H}}}_k)^{-1}$ is a positive semi-definite matrix. Based on (\ref{Psi_3}) and the following EVD with eigenvalues in decreasing order
\begin{align}
\label{DHHD}
{\bf{D}}^{-1/2}({\boldsymbol{\mathcal{H}}}^{\rm{H}}
{\boldsymbol{\mathcal{H}}})
{\bf{D}}^{-1/2}={\bf{U}}_{{\bf{M}}}
{\boldsymbol{\tilde \Lambda}}_{{\bf{M}}}{\bf{U}}_{{\bf{M}}}^{\rm{H}}
\end{align} the optimal solution of ${\bf{Q}}_k$ in (\ref{Q_3}) is further rewritten as
\begin{align}
\label{Q_final_solution}
{\bf Q}
&={\bf{D}}^{-1/2}{\bf{U}}_{{\bf{M}}}\left({\bf{I}}-
{\boldsymbol{\tilde \Lambda}}_{{\bf{M}}}^{-1}\right)^{+}
{\bf{U}}_{{\bf{M}}}^{\rm{H}}{\bf{D}}^{-1/2}
\end{align}where the symbol $[{\bf{Z}}]^+$ denotes that for the Hermitian matrix ${\bf{Z}}$ if the $i^{\rm{th}}$ largest eigenvalue is negative, if $\lambda_i({\bf{Z}})<0$ the operation $^+$ will set $\lambda_i({\bf{Z}})=0$.
It is worth noting that the formulation (\ref{Q_final_solution}) is not the exactly the optimal solution for the original optimization problem (\ref{Matrix_Opt}) as it is still a function $\alpha$.
Only if $\alpha \rightarrow 0$, (\ref{Q_final_solution}) will become to be the optimal solution of the original optimization problem (\ref{Matrix_Opt}).

Without loss of generality,  assuming the number of nonzero singular values of ${\bf{H}}$ is $K$,
the following singular value decomposition is first defined as
\begin{align}
\label{Matrix_Singular}
{\bf{H}}^{\rm{H}}{\bf{R}}_{\bf{n}}^{-1}{\bf{H}}={\bf{U}}_{\bf{H}}\left[ {\begin{array}{*{20}c}
   {{\boldsymbol{\Lambda}}_{\bf{H}}} & {{\bf{0}}}  \\
   {{\bf{0}}} & {{\bf{0}}}  \\
\end{array}} \right]{\bf{U}}_{\bf{H}}^{\rm{H}}
\end{align}where the diagonal elements of the $K\times K$ diagonal matrix ${\boldsymbol \Lambda}_{\bf{H}}$ are positive real values in decreasing order. Together with the relation $\lim_{\alpha \rightarrow 0}{\boldsymbol{\mathcal{H}}}^{\rm{H}}
{\boldsymbol{\mathcal{H}}}={\bf{H}}^{\rm{H}}{\bf{R}}_{\bf{n}}^{-1}{\bf{H}}=
[{\bf{U}}_{\bf{H}}]_{:,1:K}
{\boldsymbol{\Lambda}}_{\bf{H}}
[{\bf{U}}_{\bf{H}}]_{:,1:K}^{\rm{H}}$. When $\alpha \rightarrow 0$, for the diagonal matrix ${\boldsymbol {\tilde \Lambda}}_{\bf{M}}$ in (\ref{Q_final_solution}) only the first $K$ diagonal elements are nonzero and the remaining diagonal elements tend to be zero. Therefore, we will have the following equation
\begin{align}
\lim_{\alpha \rightarrow 0}{\bf Q}
=&{\bf{D}}^{-1/2}{\bf{U}}_{\bf{M}}\left({\bf{I}}-
{\boldsymbol {\tilde \Lambda}}_{\bf{M}}^{-1}\right)^{+}
{\bf{U}}_{\bf{M}}^{\rm{H}}{\bf{D}}^{-1/2} \nonumber \\ =&{\bf{D}}^{-1/2}[{\bf{U}}_{\bf{M}}]_{:,1:K}\left({\bf{I}}-
[{\boldsymbol {\tilde \Lambda}}_{\bf{M}}]_{1:K,1:K}^{-1}\right)^{+}
[{\bf{U}}_{\bf{M}}]_{:,1:K}^{\rm{H}}{\bf{D}}^{-1/2}.
\end{align} In addition, when $\alpha \rightarrow 0$ based on the EVD in (\ref{DHHD}) the following equality holds
\begin{align}
& \lim_{\alpha \rightarrow 0}[{\bf{U}}_{\bf{M}}]_{:,1:K}
[{\boldsymbol {\tilde \Lambda}}_{\bf{M}}]_{1:K,1:K}
[{\bf{U}}_{\bf{M}}]_{:,1:K}^{\rm{H}}
= {\bf{D}}^{-1/2}
[{\bf{U}}_{\bf{H}}]_{:,1:K}{\boldsymbol{\Lambda}}_{\bf{H}}
[{\bf{U}}_{\bf{H}}]_{:,1:K}^{\rm{H}}{\bf{D}}^{-1/2}.
\end{align} In other words, when $\alpha \rightarrow 0$ the Eigenvalue Decomposition (EVD) of ${\bf{D}}^{-1/2}
[{\bf{U}}_{\bf{H}}]_{:,1:K}{\boldsymbol{\Lambda}}_{\bf{H}}
[{\bf{U}}_{\bf{H}}]_{:,1:K}^{\rm{H}}{\bf{D}}^{-1/2}$ can be denoted as
 $[{\bf{U}}_{\bf{M}}]_{:,1:K}[{\boldsymbol {\tilde \Lambda}}_{\bf{M}}]_{1:K,1:K}
[{\bf{U}}_{\bf{M}}]_{:,1:K}^{\rm{H}}$. In summary, we have the following result.

$ \ $

\noindent \textbf{{Conclusion 2:}} The optimal ${\bf{Q}}$ of (\ref{Matrix_Opt}) has the following formulation
\begin{align}
\label{matrix_water_filling_final}
{\bf Q}
&={\bf{D}}^{-1/2}[{\bf{U}}_{\bf{M}}]_{:,1:K}
({\bf{I}}-
[{\boldsymbol {\tilde \Lambda}}_{\bf{M}}]_{1:K,1:K}^{-1})^{+}
[{\bf{U}}_{\bf{M}}]_{:,1:K}^{\rm{H}}{\bf{D}}^{-1/2}
\end{align}where the diagonal matrix $[{\boldsymbol {\tilde \Lambda}}_{\bf{M}}]_{1:K,1:K}$ and the first $K$ columns of a unitary matrix $[{\bf{U}}_{\bf{M}}]_{:,1:K}$ are computed based on the following EVD
\begin{align}
\label{EVD}
&{\bf{D}}^{-1/2}
[{\bf{U}}_{\bf{H}}]_{:,1:K}{\boldsymbol{\Lambda}}_{\bf{H}}
[{\bf{U}}_{\bf{H}}]_{:,1:K}^{\rm{H}}{\bf{D}}^{-1/2}=[{\bf{U}}_{\bf{M}}]_{:,1:K}
[{\boldsymbol {\tilde \Lambda}}_{\bf{M}}]_{1:K,1:K}
[{\bf{U}}_{\bf{M}}]_{:,1:K}^{\rm{H}}.
\end{align}

$ \ $

However the previous equation (\ref{matrix_water_filling_final}) gives the exact formula of the optimal solution. Unfortunately, it is too complicated and the variables are coupled with each other. In the following, we will proceed to simplify it. We begin with discussing a special case of ${\bf{H}}^{\rm{H}}{\bf{R}}_{\bf{n}}^{-1}{\bf{H}}$ is full rank, which is much easier.
%Following that, the general case of ${\bf{H}}^{\rm{H}}{\bf{R}}_{\bf{n}}{\bf{H}}$ is ill rank is investigated.

\noindent \textbf{The special case of
${\bf{H}}^{\rm{H}}{\bf{R}}_{\bf{n}}^{-1}{\bf{H}}$ is full rank. }

Based on (\ref{EVD}), if ${\bf{H}}^{\rm{H}}{\bf{R}}_{\bf{n}}^{-1}{\bf{H}}$ is full rank the second term in (\ref{matrix_water_filling_final}) satisfy the following relation directly
\begin{align}
\label{Water_1}
& {\bf{D}}^{-1/2}[{\bf{U}}_{\bf{M}}]_{:,1:K}
[{\boldsymbol {\tilde \Lambda}}_{\bf{M}}]_{1:K,1:K}^{-1}
[{\bf{U}}_{\bf{M}}]_{:,1:K}^{\rm{H}}{\bf{D}}^{-1/2} \nonumber \\
=&{\bf{D}}^{-1/2}{\bf{D}}^{1/2}
[{\bf{U}}_{\bf{H}}]_{:,1:K}{\boldsymbol{\Lambda}}_{\bf{H}}^{-1}
[{\bf{U}}_{\bf{H}}]_{:,1:K}^{\rm{H}}{\bf{D}}^{1/2}{\bf{D}}^{-1/2}\nonumber \\
=&[{\bf{U}}_{\bf{H}}]_{:,1:K}{\boldsymbol{\Lambda}}_{\bf{H}}^{-1}
[{\bf{U}}_{\bf{H}}]_{:,1:K}^{\rm{H}} \nonumber \\
=&({\bf{H}}^{\rm{H}}{\bf{R}}_{\bf{n}}^{-1}{\bf{H}})^{-1},
\end{align} based on which and together with the fact that in this case $[{\bf{U}}_{\bf{M}}]_{:,1:K}[{\bf{U}}_{\bf{M}}]_{:,1:K}^{\rm{H}}=
{\bf{U}}_{\bf{M}}{\bf{U}}_{\bf{M}}^{\rm{H}}={\bf{I}}$, in high SNR region the symbol $+$ can be removed and  the optimal solution of ${\bf{Q}}$ becomes to be
\begin{align}
\label{Q_1}
{\bf{Q}}={\bf{D}}^{-1}-({\bf{H}}^{\rm{H}}{\bf{R}}_{\bf{n}}^{-1}{\bf{H}})^{-1} .
\end{align}Considering ${\bf{D}}$ is diagonal and together with the power constraints, the diagonal elements of ${\bf{D}}$  can be easily solved to be
\begin{align}
\label{Computation_D}
[{\bf{D}}]_{j,j}=\frac{1}{p_k+\sum_{ j\in \psi_k }({\bf{H}}^{\rm{H}}{\bf{R}}_{\bf{n}}^{-1}{\bf{H}})^{-1}} \ \ \ j\in \psi_k.
\end{align} In the general case with ${\bf{H}}^{\rm{H}}{\bf{R}}_{\bf{n}}^{-1}{\bf{H}}$ being ill rank, the derivation of the optimal solutions becomes more challenging and this is the focus of the following section.

\section{The Proposed Solutions for the Case of Ill-Rank ${\bf{H}}^{\rm{H}}{\bf{R}}_{\bf{n}}^{-1}{\bf{H}}$ }

In this section, we are concerned on the problem how to compute the optimal solution for the general case. This problem will be solved following two different logics. The first one is an iterative solution and the second one is non-iterative solution. Specifically, iterative solutions rely on iterative computation procedure to soften the difficulty. Generally, iterative solution has a performance advantage at the expense of computational complexity. On the other hand, a non-iterative solution is given as well, which has a much clearer physical meaning. Moreover, it does not need any iterations and then has an advantage in terms of computational complexity.

For the first term on the righthand side of (\ref{matrix_water_filling_final}), based on (\ref{EVD}) $[{\bf{U}}_{\bf{M}}]_{:,1:K}$ is a function of ${\bf{D}}$. It is very challenging to formulate an explicit function of ${\bf{D}}$ to represent  $[{\bf{U}}_{\bf{M}}]_{:,1:K}$. We can only argue that $[{\bf{U}}_{\bf{M}}]_{:,1:K}$ and ${\bf{D}}$ are coupled with each other. To circumvent this difficulty, iterative algorithms are natural choices.
In the general case, in \ref{App_2} it is proved that the second term on the righthand side of (\ref{matrix_water_filling_final}) has the following property
\begin{align}
\label{second_term}
&[{\bf{U}}_{\bf{H}}]_{1:K}^{\rm{H}} {\bf{D}}^{-1/2}[{\bf{U}}_{\bf{M}}]_{:,1:K}
[{\boldsymbol {\tilde \Lambda}}_{\bf{M}}]_{1:K,1:K}^{-1}
[{\bf{U}}_{\bf{M}}]_{:,1:K}^{\rm{H}}{\bf{D}}^{-1/2}[{\bf{U}}_{\bf{H}}]_{:,1:K}={\boldsymbol{\Lambda}}_{\bf{H}}^{-1}
.
\end{align} This statement reveals a fact that no matter how transmit power varies, the term given by (\ref{second_term}) is constant.
Based on this fact in the following an iterative solution is first proposed.

\noindent \textbf{Iterative solution}

From low to moderate SNRs, for a general diagonal matrix ${\bf{D}}$ the operation of $+$ in (\ref{matrix_water_filling_final}) prohibits us from precisely analyzing the optimal solutions. Therefore, an auxiliary variable $T$ is introduced, which denotes the number of diagonal elements of $[{\boldsymbol {\tilde \Lambda}}_{\bf{M}}]_{1:K,1:K}$ larger than one, and then (\ref{matrix_water_filling_final}) becomes
\begin{align}
\label{Q_29}
{\bf Q}
&={\bf{D}}^{-1/2}[{\bf{U}}_{\bf{M}}]_{:,1:T}
[{\bf{U}}_{\bf{M}}]_{:,1:T}^{\rm{H}}{\bf{D}}^{-1/2}
-{\bf{D}}^{-1/2}[{\bf{U}}_{\bf{M}}]_{:,1:T}
[{\boldsymbol {\tilde \Lambda}}_{\bf{M}}]_{1:T,1:T}^{-1}
[{\bf{U}}_{\bf{M}}]_{:,1:T}^{\rm{H}}{\bf{D}}^{-1/2}.
\end{align} Note that the introduction of $T$ successfully removes the complicated operation $+$. Based on the previous discussions, when $K=T$ the second term in (\ref{Q_29}) is constant for the optimal ${\bf{D}}$. It is worth noting that for iterative solution, the number $T$ is unknown and when $T\not=K$
the second term in (\ref{Q_29}) cannot be guaranteed to be constant, while inspired by (\ref{second_term}) its value does not fluctuate dramatically. Therefore, in the proposed iterative solution the second term in (\ref{Q_29}) will be simply fixed at each iteration and its value is updated at the next iteration based on the value of ${\bf{D}}$ computed at the current iteration. At the $n^{\rm{th}}$ iteration, using the subscript $n$ to denote the $n^{\rm{th}}$ iteration, the signal covariance matrix ${\bf{Q}}_{n}$ is computed based on the following equation
\begin{align}
\label{iterative_solution}
{\bf Q}_n
&={\bf{D}}_n^{-1/2}
[{\bf{U}}_{{\bf{M}}_{n-1}}]_{:,1:T}
[{\bf{U}}_{{\bf{M}}_{n-1}}]_{:,1:T}^{\rm{H}}{\bf{D}}_n^{-1/2}
\nonumber \\
& \ \ \ \ \ \ \ \ \ \ \ \ \ \ \ \ \  -{\bf{D}}_{n-1}^{-1/2}[{\bf{U}}_{{\bf{M}}_{n-1}}]_{1:T}
[{\boldsymbol {\tilde \Lambda}}_{{\bf{M}}_{n-1}}]_{:,1:T,1:T}^{-1}
[{\bf{U}}_{{\bf{M}}_{n-1}}]_{:,1:T}^{\rm{H}}
{\bf{D}}_{n-1}^{-1/2}
\end{align}where $[{{\tilde \Lambda}}_{{\bf{M}}_{n-1}}]_{:,1:T,1:T}$ and $[{\bf{U}}_{{\bf{M}}_{n-1}}]_{:,1:T}$ are computed based on the following EVD
\begin{align}
\label{iterative_EVD}
&{\bf{D}}_{n-1}^{-1/2}
[{\bf{U}}_{\bf{H}}]_{:,1:K}{\boldsymbol{\Lambda}}_{\bf{H}}
[{\bf{U}}_{\bf{H}}]_{:,1:K}^{\rm{H}}{\bf{D}}_{n-1}^{-1/2}
=[{\bf{U}}_{{\bf{M}}_{n-1}}]_{:,1:K}
[{\boldsymbol {\tilde \Lambda}}_{{\bf{M}}_{n-1}}]_{1:K,1:K}
[{\bf{U}}_{{\bf{M}}_{n-1}}]_{:,1:K}^{\rm{H}}.
\end{align}In addition, based on the definition of ${\bf{D}}$ in (\ref{D}) it can be concluded that for mixed power constraint, in the same $\psi_k$ the corresponding diagonal elements of ${\bf{D}}$ are the same. Together with the fact that ${\bf{D}}$ is a diagonal matrix, ${\bf{D}}_n$ is updated based on the following equation
\begin{align}
\label{D_n}
[{\bf{D}}_n]_{j,j}=\frac{\sum_{ j\in \psi_k }[[{\bf{U}}_{{\bf{M}}_{n-1}}]_{:,1:T}
[{\bf{U}}_{{\bf{M}}_{n-1}}]_{:,1:T}^{\rm{H}}]_{j,j}}{p_k+\sum_{ j\in \psi_k }[{\bf{D}}_{n-1}^{-1/2}[{\bf{U}}_{{\bf{M}}_{n-1}}]_{1:T}
[{\boldsymbol {\tilde \Lambda}}_{{\bf{M}}_{n-1}}]_{:,1:T,1:T}^{-1}
[{\bf{U}}_{{\bf{M}}_{n-1}}]_{:,1:T}^{\rm{H}}
{\bf{D}}_{n-1}^{-1/2}]_{j,j}} \ \ \ j\in \psi_k.
\end{align}
In summary, the proposed iterative solution is given by the following pseudocode.

%\begin{algorithm}[h]
%\caption{The proposed iterative solution}
%\begin{algorithmic}[1]

%\State Initialize ${\bf{D}}_0$, e.g., ${\bf{D}}_0={\bf{I}}$.
%\Repeat

%\State{Using ${\bf{D}}_{n-1}$ computed in the preceding iteration, compute $[{\bf{U}}_{{\bf{M}}_{n-1}}]_{:,1:K}$, $
%[{\boldsymbol {\tilde \Lambda}}_{\bf{M}}]_{1:K,1:K}$ and $T$ based on (\ref{iterative_EVD}).}

%\State{Compute the diagonal matrix ${\bf{D}}_{n}$ using (\ref{D_n}) and substitute it into (\ref{iterative_solution}) to obtain %${\bf{Q}}_n$.  Reset the negative eigenvalues of the corresponding ${\bf{Q}}_n$ to be zeros. After that multiply the right-hand %and left-hand sides of ${\bf{Q}}_n$ with a diagonal matrix ${\bf{U}}_{D}$, i.e., ${\bf{U}}_{D}{\bf{Q}}_n{\bf{U}}_{D}$, to make %sure that the power constraints are satisfied.}

%\Until{The increase of the capacity is smaller than a threshold or the maximum iteration number is achieved.}

%\end{algorithmic}
%\end{algorithm}
\begin{algorithm}
%\floatname{algorithm}{Algorithm}%更改算法前缀名称
%\renewcommand{\algorithmicrequire}{\textbf{Input:}}% 更改输入名称
%\renewcommand{\algorithmicensure}{\textbf{Output:}}% 更改输出名称
\footnotesize
\caption{The proposed iterative solution}
\label{alg1}
\begin{algorithmic}[1]
    \REQUIRE ${\bf{D}}_0$, e.g., ${\bf{D}}_0={\bf{I}}$;
    \ENSURE ${\bf{Q}}$;
    \STATE {Using ${\bf{D}}_{n-1}$ computed in the preceding iteration, compute $[{\bf{U}}_{{\bf{M}}_{n-1}}]_{:,1:K}$, $
[{\boldsymbol {\tilde \Lambda}}_{\bf{M}}]_{1:K,1:K}$ and $T$ based on (\ref{iterative_EVD});}
    %\IF{$n < 0$}

         \WHILE{The increase of the capacity is larger than a threshold or the maximum iteration number is not achieved;}

         \STATE {Compute the diagonal matrix ${\bf{D}}_{n}$ using (\ref{D_n}) and substitute it into (\ref{iterative_solution}) to obtain ${\bf{Q}}_n$.  Reset the negative eigenvalues of the corresponding ${\bf{Q}}_n$ to be zeros. After that multiply the right-hand and left-hand sides of ${\bf{Q}}_n$ with a diagonal matrix ${\bf{U}}_{D}$, i.e., ${\bf{U}}_{D}{\bf{Q}}_n{\bf{U}}_{D}$, to make sure that the power constraints are satisfied;}
 %   \ELSE
  %      \STATE $X \Leftarrow x$;
 %       \STATE $N \Leftarrow n$;
 %   \ENDIF
 %   \WHILE{$N \neq 0$}
  %      \IF{$N$ is even}
  %          \STATE $X \Leftarrow X \times X$;
  %          \STATE $N \Leftarrow N / 2$;
   %     \ELSE[$N$ is odd]
   %         \STATE $y \Leftarrow y \times X$;
   %         \STATE $N \Leftarrow N - 1$;
   %     \ENDIF
    \ENDWHILE
\end{algorithmic}
\end{algorithm}
The main difference between our iterative solution and the existing work \cite{Mai2011} is that our solution still works when the rank of channel matrix is strictly smaller than both column and row numbers. Regarding the convergence property of the proposed iterative algorithm, however in the special case of ${\bf{D}}\propto {\bf{I}}$, the convergence can be proved easily \cite{Palomar2005}, for a general case with arbitrary diagonal matrix ${\bf{D}}$ it is very challenging. In the simulation part, extensive numerical simulations are exploited to show the convergence behavior of the proposed iterative solution. It is shown that the convergence property of the iterative solution is pretty good even for massive MIMO systems. Another problem is how to choose the initial values. In order to overcome these disadvantages of the iterative solution, in the following a non-iterative solution is proposed.

\noindent \textbf{Non-iterative Solution}

From (\ref{second_term}) it can be concluded that if the following equality holds,
\begin{align}
\label{constant}
&{\bf{D}}^{-1/2}[{\bf{U}}_{\bf{M}}]_{:,1:K}
[{\boldsymbol {\tilde \Lambda}}_{\bf{M}}]_{1:K,1:K}^{-1}
[{\bf{U}}_{\bf{M}}]_{:,1:K}^{\rm{H}}{\bf{D}}^{-1/2}=
[{\bf{U}}_{\bf{H}}]_{1:K}{\boldsymbol{\Lambda}}_{\bf{H}}^{-1}[{\bf{U}}_{\bf{H}}]_{:,1:K}^{\rm{H}}
,
\end{align} the equation (\ref{second_term}) can be achieved directly. It is worth noting that (\ref{constant}) is the generalized inversion of ${\bf{H}}^{\rm{H}}{\bf{R}}_{\bf{n}}^{-1}{\bf{H}}$ \cite{Hansen1998}.
We notice that in these two cases, i.e., ${\bf{H}}^{\rm{H}}{\bf{R}}_{\bf{n}}^{-1}{\bf{H}}$ is full rank ($[{\bf{U}}_{\bf{M}}]_{:,1:K}={\bf{U}}_{\bf{M}}$) or ${\bf{D}}$ proportional to identity matrix,  $[{\bf{U}}_{\bf{M}}]_{:,1:K}$ in the first term in (\ref{matrix_water_filling_final}) can replaced with $[{\bf{U}}_{\bf{H}}]_{:,1:K}$ without loss of optimality. Here, for non-iterative solution, this replacement is used for the general case. As a result, the optimal ${\bf{Q}}$ of the optimization (\ref{Matrix_Opt}) can be simplified greatly into the following simple but interesting formulation as it can be understood as a matrix version water-filling solution
\begin{align}
\label{Matrix_WF}
{\bf Q}
&=\left[{\bf{D}}^{-1/2}
[{\bf{U}}_{\bf{H}}]_{:,1:K}
[{\bf{U}}_{\bf{H}}]_{:,1:K}^{\rm{H}}{\bf{D}}^{-1/2}
-[{\bf{U}}_{\bf{H}}]_{:,1:K}{\boldsymbol{\Lambda}}_{\bf{H}}^{-1}
[{\bf{U}}_{\bf{H}}]_{:,1:K}^{\rm{H}}
\right]^{+}.
\end{align}

It is obvious that the matrix version water-filling (\ref{Matrix_WF}) includes (\ref{Q_1}) and traditional water-filling solution \cite{Telatar1999,Palomar03} as its special cases. If ${\bf{D}}$ is proportional to an identity matrix, (\ref{Matrix_WF}) will reduce to the traditional water-filling solution and this case corresponds to the transceiver designs under sum power constraint \cite{Telatar1999,Palomar03}. For the matrix version water-filling solution, ${\bf{D}}^{-1/2}
[{\bf{U}}_{\bf{H}}]_{:,1:K}
[{\bf{U}}_{\bf{H}}]_{:,1:K}^{\rm{H}}{\bf{D}}^{-1/2}$ is the matrix version water-filling level. The formulation of matrix version water-filling is just a weighted operation for $[{\bf{U}}_{\bf{H}}]_{:,1:K}
[{\bf{U}}_{\bf{H}}]_{:,1:K}^{\rm{H}}$ in matrix field \cite{XingCL2013}. It is different from the classical water-filling solutions which simply multiply $[{\bf{U}}_{\bf{H}}]_{:,1:K}
[{\bf{U}}_{\bf{H}}]_{:,1:K}^{\rm{H}}$ with a scalar. Meanwhile, $[{\bf{U}}_{\bf{H}}]_{:,1:K}{\boldsymbol{\Lambda}}_{\bf{H}}^{-1}
[{\bf{U}}_{\bf{H}}]_{:,1:K}^{\rm{H}}$ is the matrix version water bottom.

The operation $+$ is also extended from vector version to matrix version, which guarantees the positivity of the whole matrix. In other words, the matrix solution must be positive semi-definite instead a vector with each element being  nonnegative. At high SNR, the matrix water-filling level (the first term) will be much larger than the second term and thus $+$ can be simply removed, and then we have
\begin{align}
\label{Matrix_WF_1}
{\bf Q}
&={\bf{D}}^{-1/2}
[{\bf{U}}_{\bf{H}}]_{:,1:K}
[{\bf{U}}_{\bf{H}}]_{:,1:K}^{\rm{H}}{\bf{D}}^{-1/2}
-[{\bf{U}}_{\bf{H}}]_{:,1:K}{\boldsymbol{\Lambda}}_{\bf{H}}^{-1}
[{\bf{U}}_{\bf{H}}]_{:,1:K}^{\rm{H}}.
\end{align}
The solution given by (\ref{Matrix_WF_1}) is of great importance in high SNR region. In high SNR region, the computation of the diagonal matrix ${\bf{D}}$ becomes much easy. Notice that ${\bf{D}}$ is diagonal and then based on its definition in (\ref{D}) the $j^{th}$ element of ${\bf{D}}$ with $j\in \psi_k$ equals
\begin{align}
\label{Computation_D}
[{\bf{D}}]_{j,j}=\frac{\sum_{ j\in \psi_k }[[{\bf{U}}_{\bf{H}}]_{:,1:K}
[{\bf{U}}_{\bf{H}}]_{:,1:K}^{\rm{H}}]_{j,j}}{p_k+\sum_{ j\in \psi_k }[[{\bf{U}}_{\bf{H}}]_{:,1:K}{\boldsymbol{\Lambda}}_{\bf{H}}^{-1}
[{\bf{U}}_{\bf{H}}]_{:,1:K}^{\rm{H}}]_{j,j}} \ \ \ j\in \psi_k.
\end{align}
However, high SNRs are usually desired for reliable communications \cite{Gao2,Zhang2013,Zhang2014}, we still want to make the proposed solution applicable for any value of SNR.

Directly using (\ref{Computation_D}) to compute ${\bf{D}}$ cannot guarantee the positivity of the eigenvalues of ${\bf{Q}}$. Here a brute-force method is utilized.
After computing ${\bf{D}}$ based on (\ref{Computation_D}) and substituting it into (\ref{Matrix_WF_1}), the negative eigenvalues of the resulting ${\bf{Q}}$ are forced to be zeros directly. This brute-force operation will increase the diagonal elements of ${\bf{Q}}$ as some eigenvalues of ${\bf{Q}}$ increases from negative values into zeros. As a result, the power constraints will be exceeded and it means that the solution is not feasible. Then we multiply a diagonal matrix ${\bf{U}}_{\bf{D}}$ on both the righthand and lefthand sides of ${\bf{Q}}$, i.e., ${\bf{U}}_{\bf{D}}{\bf{Q}}{\bf{U}}_{\bf{D}}^{\rm{H}}
$, to make sure the power constraints satisfied. It is worth noting that the resulting new signal covariance matrix is still positive semi-definite. The diagonal matrix ${\bf{U}}_{\bf{D}}$ is defined as
\begin{align}
[{\bf{U}}_{\bf{D}}]_{j,j}=\sqrt{p_k/(\sum_{ j\in \psi_k }{[{\bf{Q}}]_{j,j}})^{-1}}.
\end{align}

We want to highlight that non-iterative algorithm is very attractive for practical implementation because several reasons. Referring to iterative algorithms, in most cases it cannot be guaranteed the global optimality even with proved convergence. Moreover, for iterative algorithms the final solutions have close relationship with initial values. Even if an iterative algorithm can be proved to converge to globally optimal solutions, the iteration numbers cannot be predicted a priori. As a result, the scare wireless resources such as hardware memory, power, etc., cannot be allocated precisely for iterative algorithms in the system design stage.

\textbf{Remark:} In this paper, perfect CSI is assumed, however in practical systems, many factors will result in imperfect CSI. As a result, the designs with imperfect CSI are of importance as they can reduce the negative effects of channel errors \cite{Liu_1,Liu_2,Liu_3,Xing}. Robust designs for MIMO systems under per-antenna power constraints are also an important research topic for future research.

\begin{figure}[!h] \centering\label{fig_1}
\includegraphics[width=3.6in]{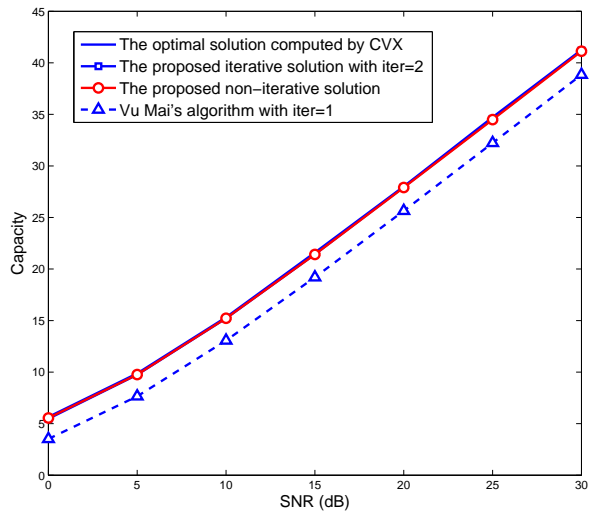}
\caption{Performance comparison between the proposed solutions and the optimal solution solved by CVX when $M=4$ and $N=8$.}
\end{figure}
\begin{figure}[!h] \centering
\includegraphics[width=3.4in]{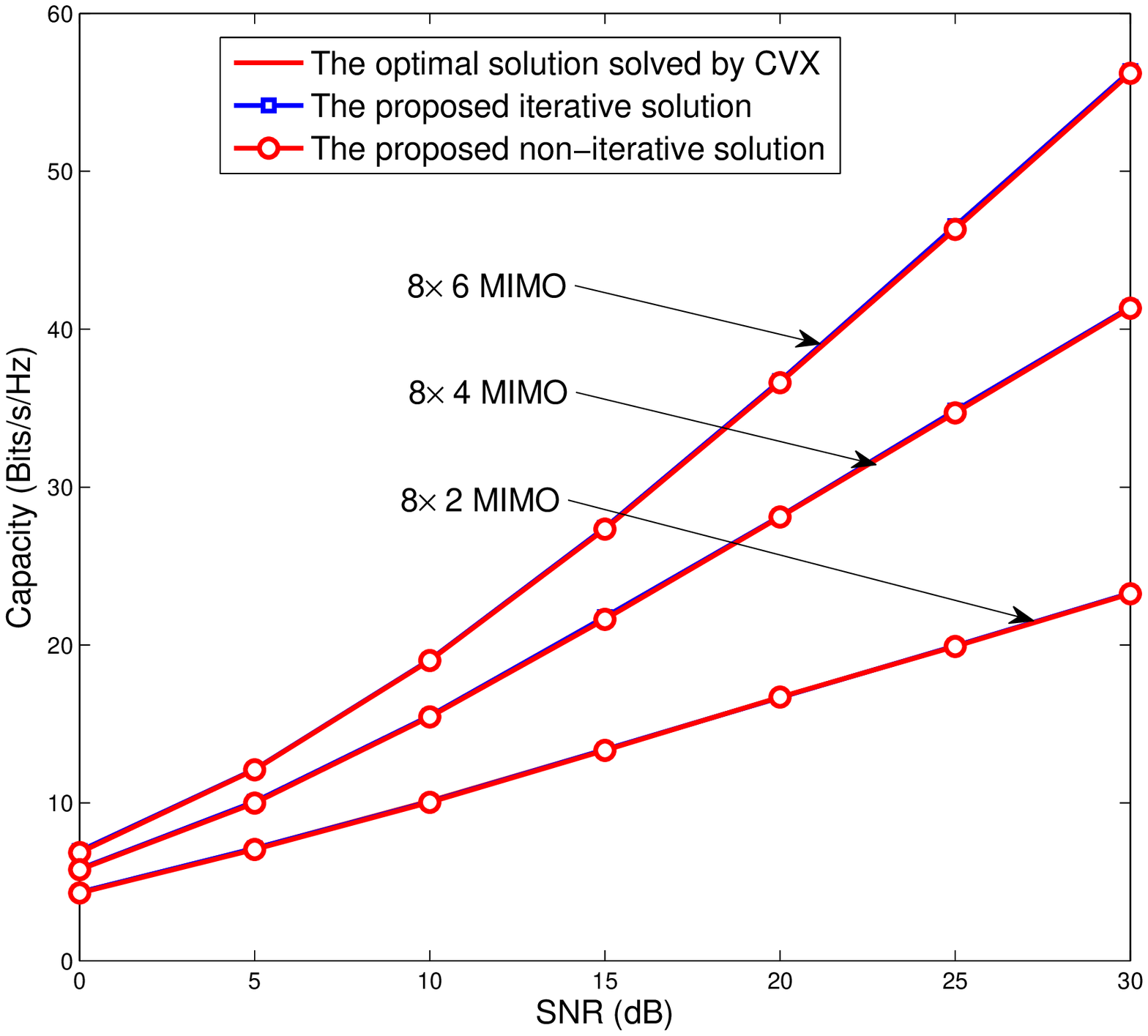}
\caption{The comparisons between the optimal solution and the proposed solutions with different settings with more transmit antennas.} \label{fig_2}
\end{figure}
\begin{figure}[!h] \centering
\includegraphics[width=3.4in]{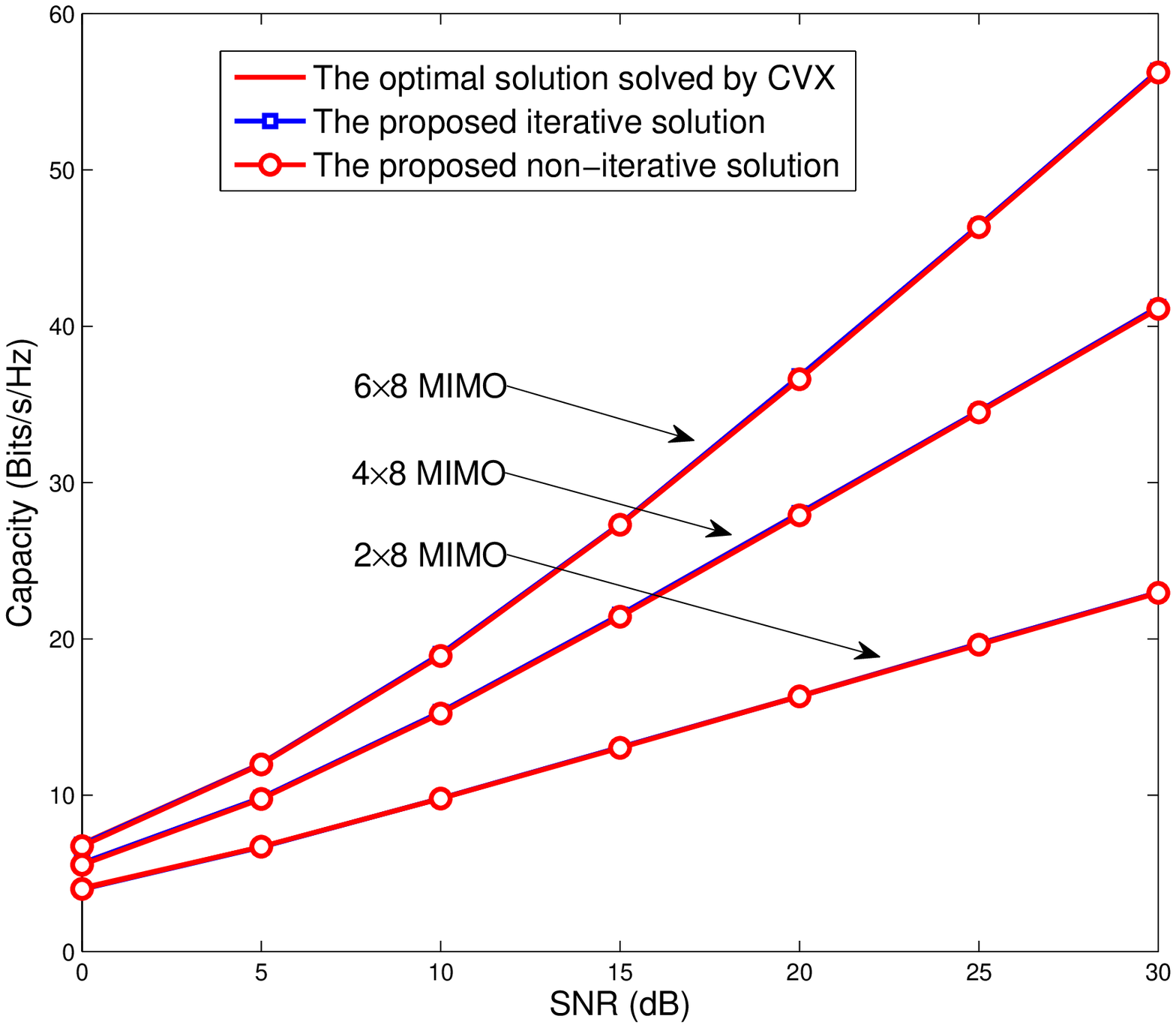}
\caption{The comparisons between the optimal solution and the proposed solutions with different settings with more receive antennas.} \label{fig_3}
\end{figure}
\begin{figure}[!h] \centering
\includegraphics[width=3.4in]{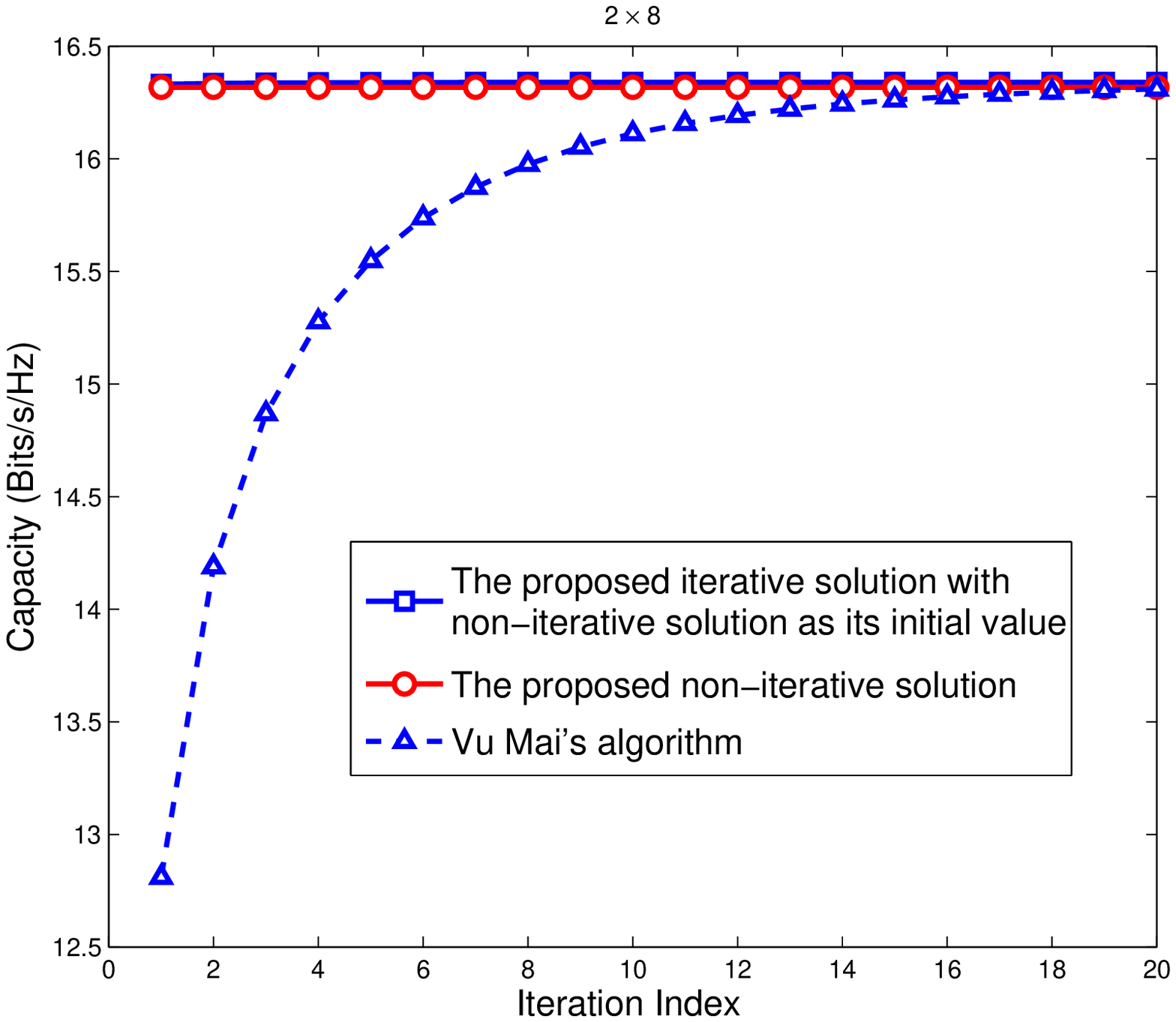}
\caption{The convergence speed of the proposed iterative solution at SNR=20dB with $M=2$ and $N=8$.} \label{fig_4}
\end{figure}

\begin{figure}[!h] \centering
\includegraphics[width=3.4in]{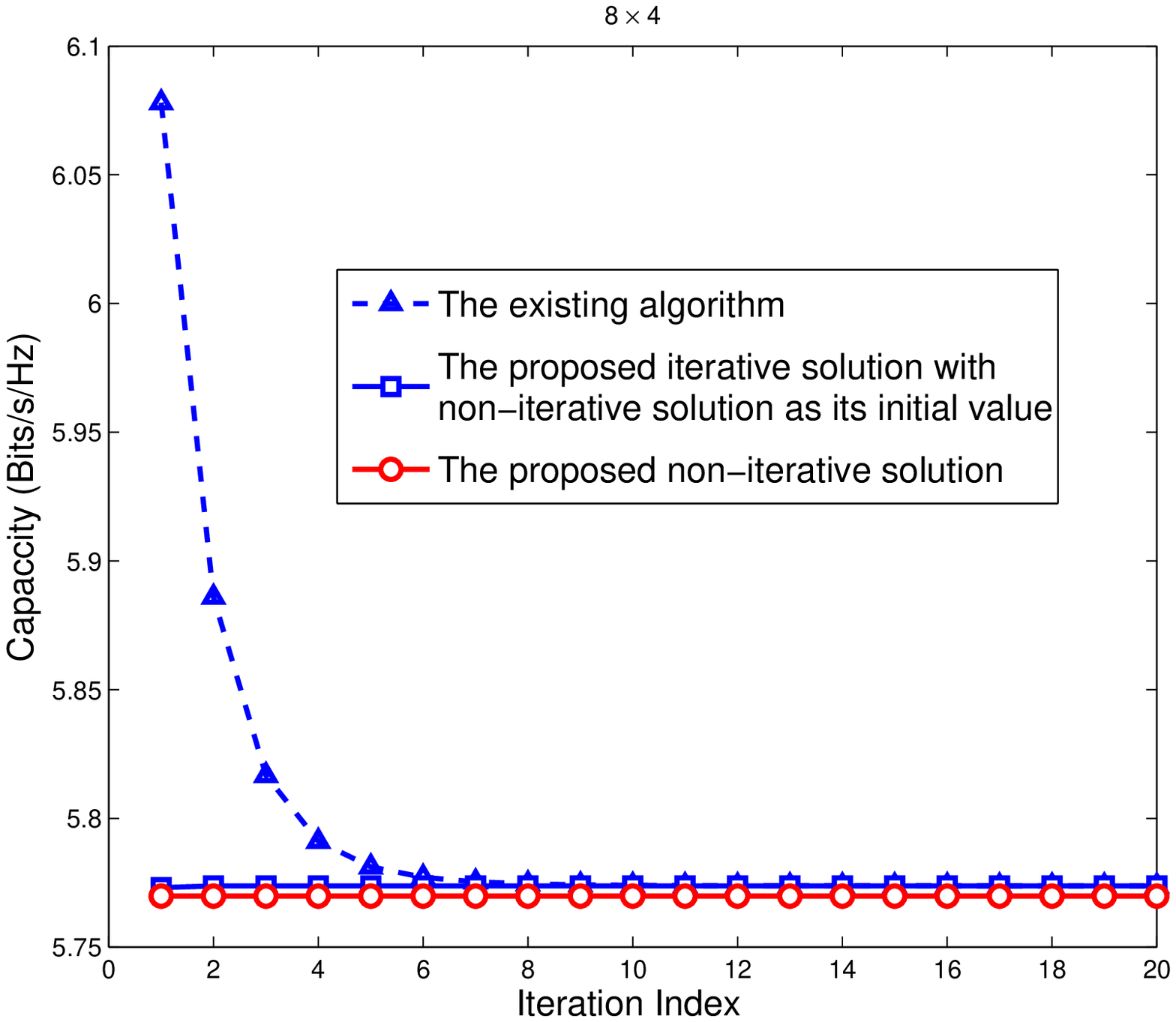}
\caption{The convergence speed of the proposed iterative solution at SNR=0dB with $M=8$ and $N=4$.} \label{fig_5}
\end{figure}

\begin{figure}[!h] \centering
\includegraphics[width=3.6in]{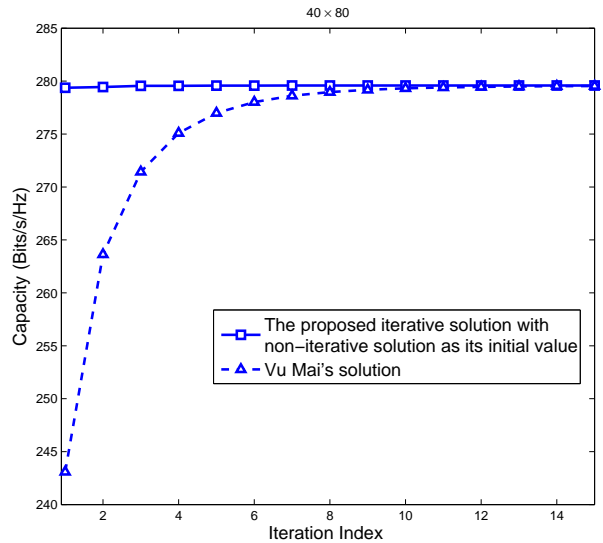}
\caption{The convergence speed of the proposed iterative solution at SNR=20dB with $M=40$ and $N=80$.} \label{fig_6}
\end{figure}
\section{Simulation Results}

In this section, our theoretical conclusions are assessed by the simulation results. In specific, a point-to-point MIMO system under per-antenna power constraint is simulated, with $N$ transmit antennas and $M$ receive antennas. Notice that as the optimization problem (\ref{Matrix_Opt}) is convex, it can be directly solved using some famous optimization software toolboxes \cite{Wiesel2008}, e.g., CVX software toolbox \cite{Grant07}. The solution of (\ref{Matrix_Opt}) computed by CVX will act as a benchmark in the following comparisons. In addition, the signal-to-noise ratio is defined as $P/\sigma_n^2$ where $P$ is total transmit power and $\sigma_n^2$ is the noise variance. To make comparisons between existing work, per-antenna power constraints are adopted in the simulation. Specifically, the power ratio from the $1^{\rm{st}}$ antenna to the $N^{\rm{th}}$ antenna is arbitrarily chosen as $N:N-1:\cdots:1$. This setting aims at making different antenna subject to significantly different power constraints. Then  per-antenna power constraint model is much more important than simple sum power constraint model. In the following figures, each point is an average of 500 channel realizations.

In Fig. 1, it is shown that the proposed iterative solution has exactly the same performance as the optimal solution solved by CVX and its convergence speed is very fast. It is worth noting that the proposed non-iterative solution also has almost the same performance as the optimal solution. In addition, without iteration the proposed non-iterative performs better than the existing algorithm in  \cite{Mai2011} via setting the iteration number of the algorithm in  \cite{Mai2011} to be 1. Furthermore, in Figs. 2 and 3, it is shown that for various simulation settings both the proposed solutions always have a pretty good performance very close to the optimal solutions solved by CVX, no matter more transmit antennas or more receive antennas. We have also tried a lot of simulation settings and the similar results can always be achieved. Due to space limitation, these results are not listed in this section.

As convergence of iterative algorithms is a critical issue, extensive simulations are performed to evaluate the convergence behavior of the proposed iterative solution. Generally speaking the proposed iterative solution enjoys a very fast convergence speed for all the simulation settings. Shown in Fig.~\ref{fig_4}, for the antenna setting with more transmit antennas, $M=6$ and $N=8$, taking the non-iterative solution as the initial value, the iterative solution converges very faster than the algorithm proposed in \cite{Mai2011}. On the other hand, the performance of the proposed non-iterative solution is satisfied. A similar result is also achieved  in the setting with more receive antennas shown in Fig. 5. It should be highlighted that in Fig. 5, the curve of the existing algorithm in \cite{Mai2011} is decreasing with iteration number. It is because in \cite{Mai2011} the iterative algorithm iteratively finds the optimal solutions satisfying the power constraints. For Vu Mai' algorithm, the iterative procedure is based on iteratively checking whether the per-antenna powers are satisfied or not. Before it converges, the solution of Vu Mai's algorithm does not satisfy the per-antenna power constraints. The reason why in Fig. 5 before converging the capacity of the solution is larger than the optimal one is in the iterative computing procedure the power constraints are not satisfied.

After that we vary the antenna numbers and even a $40 \times 80$ massive MIMO system is simulated. It is interesting that the antenna array increases, the proposed  iterative solution still enjoys a fast convergence speed. Even in a $40\times 80$ massive MIMO system, for the proposed iterative solution the convergence is achieved after a small number of iterations. It means that the proposed solution is well-suited to massive or large MIMO systems that enable high special efficiency for future communication systems.

\section{Conclusions}
In this paper, transceiver designs for MIMO systems under mixed power constraint were discussed. With mixed power constraint, some of the antennas at transmitter have their sum power constraints while the other ones are subject to per-antenna power constraints. As a result, both the transceiver designs under sum power constraint and per-antenna constraint can be considered as the special cases of the considered transceiver design. This kind of designs also have several important application scenarios e.g., C-RANs. Furthermore, the exact formula of the optimal solution has been derived. In order to compute the solution, both iterative and non-iterative solutions were proposed in this paper. The non-iterative solution has a very simple formula and can be interpreted as matrix version water-filling solution, an extension from vector domain to matrix domain. At the end the performance of the proposed solutions was assessed by the simulation results.

%%%%%%%%%%%%%%%%%%%%%%%%%%%%%%%%%%%%%%%%%%%%%%%%%%%%%%%
%%% Acknowledgements. 致谢, 非必选
%%%%%%%%%%%%%%%%%%%%%%%%%%%%%%%%%%%%%%%%%%%%%%%%%%%%%%%
%\Acknowledgements{}

%%%%%%%%%%%%%%%%%%%%%%%%%%%%%%%%%%%%%%%%%%%%%%%%%%%%%%%
%%% Supplements. 补充材料, 非必选
%%%%%%%%%%%%%%%%%%%%%%%%%%%%%%%%%%%%%%%%%%%%%%%%%%%%%%%
%\Supplements{}

%%%%%%%%%%%%%%%%%%%%%%%%%%%%%%%%%%%%%%%%%%%%%%%%%%%%%%%
%%% Reference section. 参考文献
%%% citation in the content using "some words~\cite{1,2}".
%%% ~ is needed to make the reference number is on the same line with the word before it.
%%%%%%%%%%%%%%%%%%%%%%%%%%%%%%%%%%%%%%%%%%%%%%%%%%%%%%%

%%%%%%%%%%%%%%%%%%%%%%%%%%%%%%%%%%%%%%%%%%%%%%%%%%%%%%%
%%% Appendix sections. 附录章节, 非必选
%%%%%%%%%%%%%%%%%%%%%%%%%%%%%%%%%%%%%%%%%%%%%%%%%%%%%%%
\begin{appendix}
%\section{Name}
\section{Proof of Conclusion 1}
\label{App_1}
\noindent \textsl{Proof:} Because $({\bf{I}}-{\bf{D}}^{-1/2}{\boldsymbol \Psi}{\bf{D}}^{-1/2})^{-1}={\bf{I}}+{\bf{D}}^{1/2}{\boldsymbol \Pi}{\bf{D}}^{1/2}$, denoting ${\boldsymbol \Phi}={\bf{D}}^{1/2}{\boldsymbol \Pi}{\bf{D}}^{1/2}$ we have
\begin{align}
\label{app_matrix}
{\bf{D}}^{-1/2}{\boldsymbol \Psi}{\bf{D}}^{-1/2}={\bf{I}}-({\bf{I}}+{\boldsymbol \Phi})^{-1},
\end{align} from which we can easily draw a conclusion that for a positive semi-definite matrix ${\bf{M}}$,
if ${\rm{Tr}}({\bf{M}}{\boldsymbol \Phi})=0$ we will have ${\rm{Tr}}({\bf{M}}{\bf{D}}^{-1/2}{\boldsymbol \Psi}{\bf{D}}^{-1/2})=0$.
Together with the fact that
\begin{align}
{\rm{Tr}}[({\bf{I}} -{\bf{D}}^{1/2}({\boldsymbol{\mathcal{H}}}^{\rm{H}}{\boldsymbol{\mathcal{H}}})^{-1}{\bf{D}}^{1/2}
+{\boldsymbol \Phi}){\boldsymbol \Phi}]=0,
 \end{align}
 we straightforwardly have ${\rm{Tr}}[({\bf{I}} -{\bf{D}}^{1/2}({\boldsymbol{\mathcal{H}}}^{\rm{H}}{\boldsymbol{\mathcal{H}}})^{-1}{\bf{D}}^{1/2}+{\boldsymbol \Phi}){\bf{D}}^{-1/2}{\boldsymbol \Psi}{\bf{D}}^{-1/2}]=0$ based on which it is obvious that
\begin{align}
&{\rm{Tr}}[\underbrace{{\bf{D}}^{-1/2}[{\bf{I}} -{\bf{D}}^{1/2}({\boldsymbol{\mathcal{H}}}^{\rm{H}}{\boldsymbol{\mathcal{H}}})^{-1}{\bf{D}}^{1/2} +{\boldsymbol \Phi}]{\bf{D}}^{-1/2}}_{={\bf{Q}}}{\boldsymbol \Psi}]=0. \nonumber
\end{align}
%It should be highlighted that the lefthand side of the previous equation is exactly ${\bf{Q}}$, i.e.,
%\begin{align}
%{\bf Q}=&{\bf{D}}^{-1/2}[{\bf{I}} -{\bf{D}}^{1/2}({\bf{H}}^{\rm{H}}{\bf{H}})^{-1}{\bf{D}}^{1/2} +{\boldsymbol \Phi}]{\bf{D}}^{-1/2}.
%\end{align}
Finally, it can be concluded that
\begin{align}
&{\rm{Tr}}[({\bf{I}} -{\bf{D}}^{1/2}({\boldsymbol{\mathcal{H}}}^{\rm{H}}{\boldsymbol{\mathcal{H}}})^{-1}{\bf{D}}^{1/2}+{\boldsymbol \Phi}){\boldsymbol \Phi}]=0 \rightarrow {\rm{Tr}}({\bf{Q}}{\boldsymbol \Psi})=0. \nonumber
\end{align} $\blacksquare$

\section{The proof of (\ref{second_term})}
\label{App_2}
\textsl{Proof:} Based on the EVD defined in (\ref{EVD}), $[{\boldsymbol {\tilde \Lambda}}_{\bf{M}}]_{1:K,1:K}^{-1}$ is solved to be
\begin{align}
[{\boldsymbol {\tilde \Lambda}}_{\bf{M}}]_{1:K,1:K}^{-1}=([{\bf{U}}_{\bf{H}}]_{:,1:K}^{\rm{H}}{\bf{D}}^{-1/2}
[{\bf{U}}_{\bf{M}}]_{:,1:K})^{-1}{\boldsymbol{\Lambda}}_{\bf{H}}^{-1}
([{\bf{U}}_{\bf{M}}]_{:,1:K}^{\rm{H}}{\bf{D}}^{-1/2}
[{\bf{U}}_{\bf{H}}]_{:,1:K})^{-1}
\end{align}based on which the second term of (\ref{matrix_water_filling_final}) equals
\begin{align}
&{\bf{D}}^{-1/2}[{\bf{U}}_{\bf{M}}]_{:,1:K}
[{\boldsymbol {\tilde \Lambda}}_{\bf{M}}]_{1:K,1:K}^{-1}
[{\bf{U}}_{\bf{M}}]_{:,1:K}^{\rm{H}}{\bf{D}}^{-1/2}\nonumber \\=&{\bf{D}}^{-1/2}[{\bf{U}}_{\bf{M}}]_{:,1:K}
([{\bf{U}}_{\bf{H}}]_{:,1:K}^{\rm{H}}{\bf{D}}^{-1/2}
[{\bf{U}}_{\bf{M}}]_{:,1:K})^{-1}{\boldsymbol{\Lambda}}_{\bf{H}}^{-1} %\nonumber \\
%& \ \ \ \ \ \ \ \ \ \ \ \ \ \ \ \ \ \ \ \ \ \ \ \ \ \times
([{\bf{U}}_{\bf{M}}]_{:,1:K}^{\rm{H}}{\bf{D}}^{-1/2}
[{\bf{U}}_{\bf{H}}]_{:,1:K})^{-1}
[{\bf{U}}_{\bf{M}}]_{:,1:K}^{\rm{H}}{\bf{D}}^{-1/2}.
\end{align} To further simply the above complicated formula we multiply $[{\bf{U}}_{\bf{H}}]_{:,1:K}^{\rm{H}}$ and $[{\bf{U}}_{\bf{H}}]_{:,1:K}$ on the left and right sides and then we have a much simpler form given as
\begin{align}
&[{\bf{U}}_{\bf{H}}]_{:,1:K}^{\rm{H}}{\bf{D}}^{-1/2}[{\bf{U}}_{\bf{M}}]_{:,1:K}
[{\boldsymbol {\tilde \Lambda}}_{\bf{M}}]_{1:K,1:K}^{-1}
[{\bf{U}}_{\bf{M}}]_{:,1:K}^{\rm{H}}{\bf{D}}^{-1/2}[{\bf{U}}_{\bf{H}}]_{:,1:K}\nonumber \\
=&[{\bf{U}}_{\bf{H}}]_{:,1:K}^{\rm{H}}{\bf{D}}^{-1/2}[{\bf{U}}_{\bf{M}}]_{:,1:K}
([{\bf{U}}_{\bf{H}}]_{:,1:K}^{\rm{H}}{\bf{D}}^{-1/2}
[{\bf{U}}_{\bf{M}}]_{:,1:K})^{-1}{\boldsymbol{\Lambda}}_{\bf{H}}^{-1} %\nonumber \\
%& \ \ \ \ \ \ \ \ \ \ \ \ \ \ \ \ \ \ \ \ \ \ \ \ \ \times
([{\bf{U}}_{\bf{M}}]_{:,1:K}^{\rm{H}}{\bf{D}}^{-1/2}
[{\bf{U}}_{\bf{H}}]_{:,1:K})^{-1}
[{\bf{U}}_{\bf{M}}]_{:,1:K}^{\rm{H}}{\bf{D}}^{-1/2} [{\bf{U}}_{\bf{H}}]_{:,1:K} \nonumber \\
=&{\boldsymbol{\Lambda}}_{\bf{H}}^{-1}.
\end{align}
\noindent $\blacksquare$
\end{appendix}

\end{document}